\documentclass[12pt]{article}
 
\newif\ifpdf
\ifx\pdfoutput\undefined
\pdffalse 
\else
\pdfoutput=1 
\pdftrue
\fi

\ifpdf
\usepackage[pdftex]{graphicx}
\else
\usepackage{graphicx}
\fi

\usepackage{setspace}
\title{
{\normalsize \hfill SPIN-2002/40}\\
${}$\\ 
A discrete history of the Lorentzian path integral}
\author{R. Loll\\
${}$\\
{\small Institute for Theoretical Physics, Utrecht University,}\\
{\small Leuvenlaan 4, NL-3584 CE Utrecht}}

\begin{document}

\ifpdf
\DeclareGraphicsExtensions{.pdf, .jpg, .tif}
\else
\DeclareGraphicsExtensions{.eps, .jpg}
\fi

\maketitle
\abstract{In these lecture notes,
I describe the motivation behind a recent
formulation of a non-perturbative gravitational path integral for
Lorentzian (instead of the usual Euclidean) space-times,
and give a pedagogical introduction to its main features.
At the regularized, discrete level this approach
solves the problems of (i) having a well-defined Wick rotation,
(ii) possessing a coordinate-invariant cutoff, and (iii) leading to
{\it convergent} sums over geometries.
Although little is known as yet about the
existence and nature of an underlying continuum
theory of quantum gravity in four dimensions, there are already a 
number of
beautiful results in $d=2$ and $d=3$ where continuum limits
have been found. They include an explicit example of the
inequivalence of the Euclidean and Lorentzian path integrals, a
non-perturbative mechanism for the
cancellation of the conformal factor, and the discovery
that causality can act as an effective regulator of quantum
geometry. 


\section{Introduction}\label{Why quantum gravity?}

The desire to understand the quantum physics of the gravitational
interactions lies at the root of many recent developments
in theoretical high-energy physics.
By {\it quantum gravity} I will mean a consistent fundamental quantum
description of space-time geometry (with or without matter) whose 
classical limit is general relativity. Among the possible ramifications
of such a theory are a model for the structure of space-time near
the Planck scale, a consistent calculational scheme to compute
gravitational effects at all energies, a description of (quantum)
geometry near space-time singularities and a non-perturbative quantum
description of four-dimensional black holes. It might also help us in
understanding cosmological issues about the beginning (and end?)
of our universe, although it should be said that some questions (for 
example, that of the ``initial conditions") are likely to remain outside 
the scope of any physical theory. 

From what we know about the quantum dynamics of the other
fundamental interactions it seems eminently plausible that 
also the gravitational excitations should at very short scales be
governed by quantum laws, so why have we so far not been
able to determine what they are? -- One obvious obstacle is the
difficulty in finding any direct or indirect evidence for quantum
gravitational effects, be they experimental or observational,
which could provide a feedback for model-building. 
A theoretical complication is that the outstanding problems mentioned 
above require a non-perturbative treatment; it is not sufficient to
know the first few terms of a perturbation series. This is true for
both conventional perturbative path integral expansions of 
gravity or supergravity\footnote{Of course, we already 
know that in these cases a quantization based on a decomposition
$g_{\mu\nu}(x)=\eta^{\rm Mink}_{\mu\nu} +\sqrt{G_N}\ h_{\mu\nu}(x)$,
for a linear spin-2 perturbation around Minkowski space
leads to a non-renormalizable theory.} and a perturbative expansion 
in the string coupling in the case of unified approaches. One
avenue to take is to search for a {\it non-perturbative} 
definition of such a theory, where the initial input
of any fixed ``background metric'' is inessential (or even
undesirable), and where ``space-time" is determined
{\it dynamically}. Whether or not such an approach necessarily
requires the inclusion of higher dimensions and 
fundamental supersymmetry is currently unknown. 
As we will see in the course of these lecture notes, it is perfectly
conceivable that one can do without. 

Such a non-perturbative viewpoint is very much in line with
how one proceeds in classical general relativity, where a metric space-time
$(M,g_{\mu\nu})$ (+matter) emerges only as a {\it solution} to the
Einstein equations
\begin{equation}
R_{\mu\nu}[g]-\frac{1}{2}g_{\mu\nu} R[g] +\Lambda g_{\mu\nu}=
-8\pi G_N T_{\mu\nu}[\Phi],
\label{einstein}
\end{equation}
which define the classical dynamics on the space ${\cal M}(M)$,
the space of all metrics on a given differentiable manifold $M$.
The analogous question I want to address in the quantum
theory is 

\vspace{.6cm}
\parbox{13.4cm}{Can we obtain ``quantum space-time" as a 
solution to a set of non-perturbative quantum equations of
motion on a suitable quantum analogue of ${\cal M}(M)$ or
rather, of the space of geometries, Geom$(M):={\cal M}(M)/{\rm
Diff}(M)$?}
\vspace{.6cm}

\noindent This is not a completely straightforward task.
Whichever way we want to proceed non-perturbatively, if we give 
up the privileged role of a flat, Minkowskian background
space-time on which the quantization is to take place, we also
have to abandon the central role usually played by the
Poincar\'e group, and with it most standard quantum field-theoretic
tools for regularization and renormalization. If one
works in a continuum metric formulation of gravity, the 
symmetry group of the Einstein action is instead the group Diff(M) 
of diffeomorphisms on $M$, which in terms of local charts are 
simply the smooth invertible coordinate transformations $x^\mu\mapsto
y^\mu(x^\mu)$.\footnote{One should not get confused here by
the fact that in gauge formulations of gravity which work with
vierbeins $e_\mu^a$ instead of the metric tensor $g_{\mu\nu}$,
one has an additional local invariance under SO(3,1)-frame 
rotations, ie. elements of the Lorentz group, in addition to 
diffeomorphism invariance. Nevertheless, this formulation is
still not invariant under {\it global} Lorentz- or Poincar\'e
transformations.}

I will in the following describe a particular path integral approach 
to quantum gravity, which is non-perturbative from the outset
in the sense of being defined on the ``space of all geometries"
(to be defined later), without distinguishing any background metric
structure (see also \cite{ambjorn,loll} for related reviews). 
This is closely related in spirit with the canonical
approach of loop quantum gravity \cite{rovelli} and its more recent 
incarnations using so-called spin networks \cite{thiemann,oriti},
although there are significant differences in methodology and 
attitude. ``Non-perturbative" means in a
covariant context that the path sum or integral will have to be
performed explicitly, and not just evaluated around its
stationary points, which can only be achieved in an appropriate
regularization. The method I will employ uses a discrete
lattice regularization as an intermediate step in the construction
of the quantum theory. However, unlike in lattice QCD, the lattice 
and its geometric properties will not be part of a static background 
structure, but dynamical quantities, as befits a theory of {\it
quantum geometry}.

\section{Quantum gravity from dynamical triangulations}\label{qua}

In this section I will explain how one may construct a theory
of quantum gravity from a non-perturbative path integral, 
and what logic has led my collaborators and me to consider
the method of Lorentzian dynamical triangulations to achieve
this. The method is minimal in the sense of employing 
standard tools from quantum field theory and the theory of
critical phenomena and adapting them to the case of
{\it generally covariant systems}, without invoking any 
symmetries beyond those of the classical theory. At an
intermediate stage of the construction, we use a regularization
in terms of simplicial ``Regge geometries", that is, piecewise
linear manifolds. In this approach, ``computing the path integral" 
amounts to a conceptually simple and geometrically transparent 
``counting of geometries", with additional weight factors 
which are determined by the Einstein action. 
This is done first of all at a regularized level. 
Subsequently, one searches for interesting
continuum limits of these discrete models which are
possible candidates for theories of quantum gravity, a
step that will always involve a renormalization.
From the point of view of statistical mechanics, one may
think of Lorentzian dynamical triangulations as a new class
of statistical models of Lorentzian random surfaces in various
dimensions, whose building blocks are flat simplices 
which carry a ``time arrow", and whose dynamics is entirely
governed by their intrinsic geometric properties.

Before describing the details of the construction, it may be
helpful to recall the path integral representation for a 
(one-dimensional)
non-relativistic particle \cite{pibook}. 
The time evolution of the particle's
wave function $\psi$ may be described by the integral equation
\begin{equation}
\psi (x'',t'')=\int_{\bf R} G(x'',x';t'',t')\psi (x',t'),
\label{nonrel}
\end{equation}
where the propagator or Feynman kernel $G$ is defined
through a limiting procedure,
\begin{equation}
G(x'',x';t'',t')=\lim_{\epsilon\rightarrow 0} A^{-N} \prod_{k=1}^{N-1}
\int dx_k\ {\rm e}^{i\sum_{j=0}^{N-1} \epsilon 
L(x_{j+1},(x_{j+1}-x_j)/\epsilon )}.
\label{partpi}
\end{equation}
The time interval $t''-t'$ has been discretized into $N$ steps
of length $\epsilon =(t''-t')/N$, and the right-hand side of 
(\ref{partpi}) represents
an integral over all piecewise linear paths $x(t)$ of a ``virtual'' particle 
propagating from $x'$ to $x''$, illustrated in Fig.\ref{particle}. 

\begin{figure}[t]
\vspace{0.5cm}
\centerline{\scalebox{0.6}{\rotatebox{0}
{\includegraphics{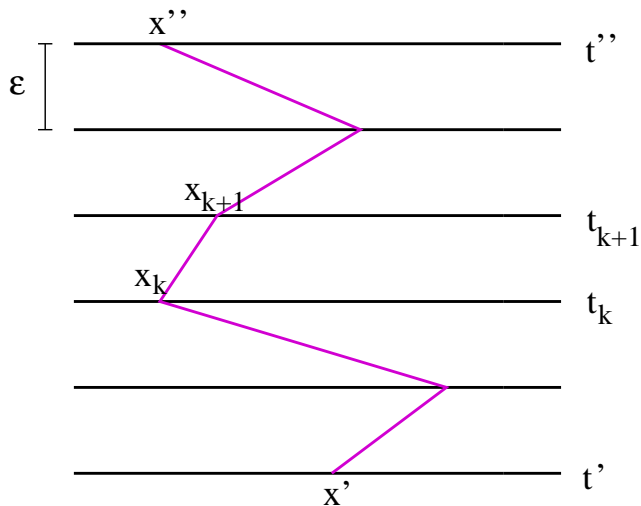}}}}
\caption[particle]{A piecewise linear particle path
contributing to the discrete Feynman propagator.}
\vspace{0.5cm}
\label{particle}
\end{figure}

The prefactor $A^{-N}$ is a normalization
and $L$ denotes the Lagrange function of the particle.
Knowing the propagator $G$ is tantamount to having
solved the quantum dynamics. This is the simplest instance of a
{\it path integral}, and is often written schematically as
\begin{equation}
G(x',t';x'',t'')=\int {\cal D}x(t)\ {\rm e}^{iS[x(t)]},
\label{feynman}
\end{equation}
where ${\cal D}x(t)$ is a functional measure on the
``space of all paths'', and the exponential weight
depends on the classical action $S[x(t)]$ of a path. Recall also that
this procedure can be defined in a mathematically
clean way if we Wick-rotate the time variable $t$ to
imaginary values $t\mapsto \tau=it$, thereby making all integrals
real \cite{reedsimon}.

Can a similar strategy work for the case of Einstein gravity? As
an analogue of the particle's position we can take the geometry
$[g_{ij}(x)]$ (ie. an equivalence class of spatial metrics) of a
constant-time slice. Can one then define a gravitational propagator
\begin{equation}
G([g_{ij}'],[g_{ij}''])=\int_{\rm Geom(M)} {\cal D}[g_{\mu\nu}]\ 
{\rm e}^{i S^{\rm Einstein} [g_{\mu\nu}]}
\label{pigrav}
\end{equation}
from an initial geometry $[g']$ to a final geometry $[g'']$
(Fig.\ref{gravprop}) as a limit of some discrete construction 
analogous to that of the
non-relativistic particle (\ref{partpi})?
And crucially, what would be a suitable class of  ``paths", that is,
space-times $[g_{\mu\nu}]$ to sum over?

\begin{figure}[h]
\vspace{0.5cm}
\centerline{\scalebox{0.5}{\rotatebox{0}
{\includegraphics{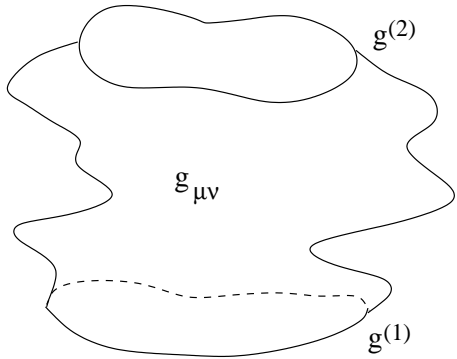}}}}
\caption[gravprop]{The time-honoured way \cite{hawking} of 
illustrating the
gravitational path integral as the propagator from an initial to a final spatial
boundary geometry.}
\vspace{0.5cm}
\label{gravprop}
\end{figure}

Setting aside the question of the physical meaning of an expression
like (\ref{pigrav}), gravitational path integrals in the continuum are
extremely ill-defined. Clearly, {\it defining} a fundamental theory of
quantum gravity via a
perturbation series in the gravitational coupling does not work
because of its perturbative non-renormalizability. 
So, is there a chance we might simply be able to  {\it do}
the integration $\int{\cal D}[g_{\mu\nu}]$ in a meaningful way?
Firstly, there is no obvious way to parametrize ``geometries'',
which means that in practice one always has to start with
gauge-{\it co}variant fields, and gauge-fix. Unfortunately,
this gives rise to Faddeev-Popov determinants whose
non-perturbative evaluation is exceedingly difficult. A similar
problem already applies to the action itself, which is by no
means quadratic, no matter what we choose as our basic fields.
How then can the integration over exp($iS$) possibly be performed? 
Part of the problem is clearly also the complex nature of this integrand,
with no obvious choice of a Wick rotation in the context of a theory
with fluctuating geometric degrees of freedom. Secondly, since we are
dealing with a field theory, some kind of regularization will be
necessary, and the challenge here is to find a procedure that
does not violate diffeomorphism-invariance.

In brief, the strategy I will be following starts from a regularized version
of the space Geom($M$) of all geometries. A regularized path
integral $G(a)$ can be defined which depends on an ultraviolet
cutoff $a$ and is {\it convergent} in a non-trivial region of the
space of coupling constants. Taking the continuum limit 
corresponds to letting $a\rightarrow 0$. The resulting continuum
theory -- if it can be shown to exist -- is then investigated with
regard to its geometric properties and in particular its semiclassical
limit. 

\section{Brief summary of discrete gravitational path integrals}\label{brief}

Trying to construct
non-perturbative path integrals for gravity from sums over discretized
geometries is not a new idea. The approach of {\it Lorentzian
dynamical triangulations} draws from older work in this area, 
but differs from it in several significant aspects as we shall see in
due course. 

Inspired by the successes of lattice gauge theory, attempts to describe
quantum gravity by similar methods have been popular on and off
since the late 70's. Initially the emphasis was on gauge-theoretic,
first-order formulations of gravity, usually based on (compactified
versions of) the Lorentz group, followed in the 80's by ``quantum
Regge calculus", an attempt to represent the gravitational path
integral as an integral over certain piecewise linear geometries
(see \cite{reggereviews} and references therein),
which had first made an appearance in approximate descriptions
of {\it classical} solutions of the Einstein equations. A variant of
this approach by the name of ``dynamical triangulation(s)" attracted
a lot of interest during the 90's, partly because it had
proved a powerful tool in describing two-dimensional quantum
gravity (see the textbook \cite{quantumgeometry} and
lecture notes \cite{iceland} for more details). 

The problem is that none of these attempts have so far come
up with convincing evidence for the existence of an underlying
continuum theory of {\it four}-dimensional quantum gravity.
This conclusion is drawn largely on the basis of numerical 
simulations, so it is by no means water-tight, although one
can make an argument that the ``symptoms'' of failure
are related in the various approaches \cite{mylivrev}. What goes wrong
generically seems to be a dominance in the continuum limit of
highly degenerate geometries, whose precise form depends on
the approach chosen. One would of course expect that non-smooth
geometries play a decisive role, in the same way as it can be
shown in the particle case that the support of the measure
in the continuum limit is on a set of nowhere differentiable
paths. However, what seems to happen in the case of 
the path integral for four-geometries is
that the structures obtained are {\it too} wild, in the
sense of not generating, even at coarse-grained scales, an
effective geometry whose dimension is anywhere near four.

The schematic phase diagram of Euclidean dynamical triangulations 
shown in Fig.\ref{phaseeu} gives an example of what can happen. The picture 
turns out to
be essentially the same in both three and four dimensions:
the model possesses infinite-volume limits everywhere along the 
critical line $k_3^{\rm crit}(k_0)$, which fixes the bare
cosmological constant as a function of the inverse Newton
constant $k_0\sim G_N^{-1}$. Along this line, there
is a critical point $k_0^{\rm crit}$ (which we now know to be of first
order in $d=3,4$) below which geometries generically have a very large
effective or Hausdorff dimension. (In terms of geometry, this means
that there are a few vertices at
which the entire space-time ``condenses'' in the sense that
almost every other vertex in the simplicial space-time is
about one link-distance away from them.) Above $k_0^{\rm crit}$
we find the opposite phenomenon of ``polymerization'':
a typical element contributing to the state sum is a thin
branched polymer, with one or more dimensions ``curled up''
(an image familiar to string theorists!) such that its effective
dimension is around two. 

\begin{figure}[h]
\vspace{0.5cm}
\centerline{\scalebox{0.6}{\rotatebox{0}
{\includegraphics{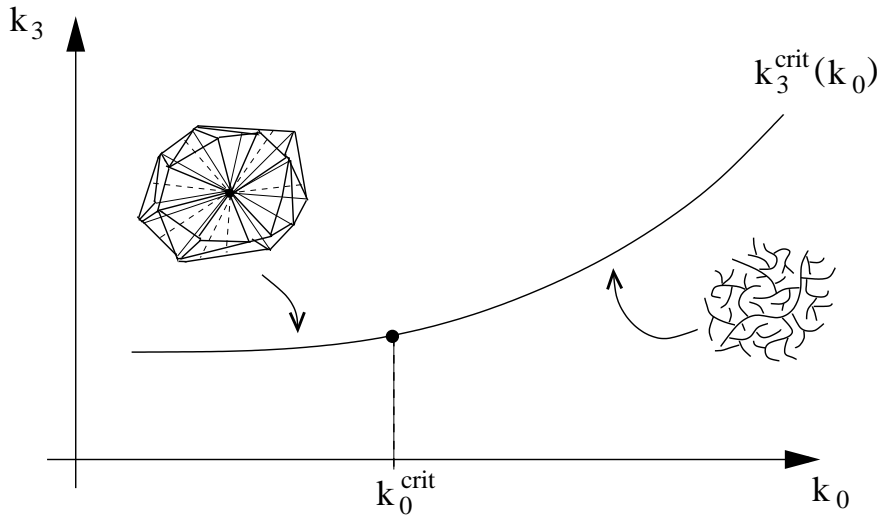}}}}
\caption[phaseeu]{The phase diagram of three- and four-dimensional 
Euclidean dynamical triangulations.}
\vspace{0.5cm}
\label{phaseeu}
\end{figure}

{\it Why} this happens was, at least until recently, less clear,
although it has sometimes been related to the so-called
conformal-factor problem. This problem has to do with the fact that
the gravitational action is unbounded below, causing
potential havoc in Euclidean versions of the path integral.
This will be discussed in more
detail below in Sec.\ \ref{three}, but it does lead directly to the
next point. Namely, what all the above-mentioned approaches
have in common is that they work from the outset with
{\it Euclidean} geometries, and associated Boltzmann-type
weights exp($-S^{\rm eu}$) in the path integral. In other
words, they integrate over ``space-times'' which know nothing
about time, light cones and causality. This is done mainly for
technical reasons, since it is difficult to set up simulations
with complex weights and since until recently a suitable Wick
rotation was not known.

``Lorentzian dynamical triangulations'', first proposed in \cite{al}
and further elaborated in \cite{ajl1,d3d4} tries to establish a logical
connection between the fact that non-perturbative path
integrals were constructed for Euclidean instead of Lorentzian
geometries and their apparent failure to lead to an interesting
continuum theory. Is it conceivable that we can kill two birds
with one stone, ie. cure the problem of degenerate quantum
geometry by taking a path integral over geometries with
a physical, Lorentzian signature? Remarkably, this is indeed
what happens in the quantum gravity theories in $d<4$ which have
already been studied extensively. The way in which Lorentzian
dynamical triangulations overcome the problems mentioned
above is the subject of the Sec.\ \ref{lorentz}.

\section{Geometry from simplices}\label{simplices}

The use of simplicial methods in general relativity goes back to
the pioneering work of Regge \cite{regge}. In classical applications
one tries to approximate a classical space-time geometry by
a triangulation, that is, a piecewise linear space obtained by
gluing together flat simplicial building blocks, which in dimension
$d$ are $d$-dimensional generalizations of triangles. By ``flat" 
I mean that they are isometric to a subspace of
$d$-dimensional Euclidean or Minkowski space. We will only
be interested in gluings leading to genuine manifolds,
which therefore look locally like an $R^d$. A nice feature of such
simplicial manifolds is that their geometric properties are
completely described by the discrete set $\{ l_i^2\}$ of the squared
lengths of their edges. Note that this amounts to a description of
geometry {\it without the use of coordinates}. There is nothing to
prevent us from re-introducing coordinate patches covering the
piecewise linear manifold, for example, on each individual
simplex, with suitable transition functions between patches.
In such a coordinate system the metric tensor will then assume a
definite form. However, for the purposes of formulating the path
integral we will not be interested in doing this, but rather work
with the edge lengths, which constitute a direct, regularized
parametrization of the space Geom($M$) of geometries. 

\begin{figure}[t]
\vspace{0.5cm}
\centerline{\scalebox{0.6}{\rotatebox{0}{\includegraphics{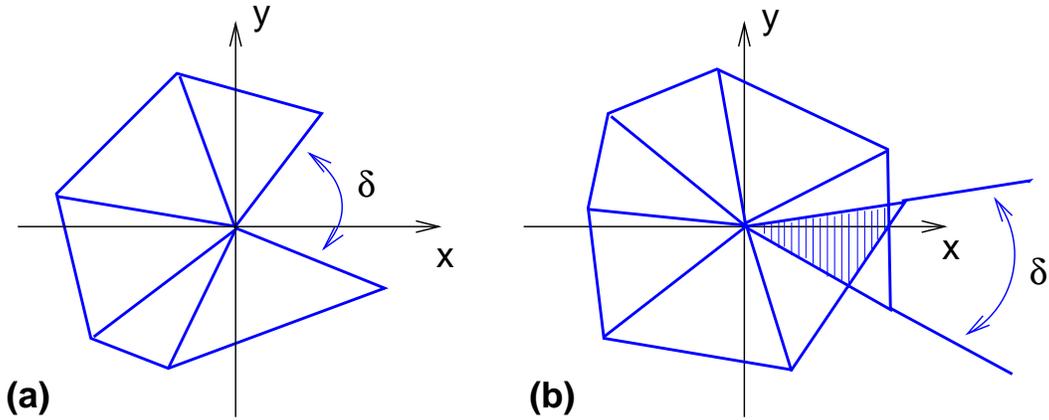}}}}
\vspace{0.5cm}
\caption[euangle]{Positive (a) and negative (b) 
space-like deficit angles $\delta$.}
\label{euangle}
\end{figure}

How precisely is the intrinsic geometry of a simplicial space,
most importantly, its curvature, encoded in its edge lengths? 
A useful example to keep in mind is the case of dimension two,
which can easily be visualized. A 2d piecewise linear space
is a triangulation, and its scalar curvature $R(x)$ coincides with
the so-called Gaussian curvature. One way of measuring this
curvature is by parallel-transporting a vector around closed
curves in the manifold. In our piecewise-flat manifold such a vector
will always return to its original orientation {\it unless} it has
surrounded lattice vertices $v$ at which the surrounding angles
did not add up to $2\pi$, but $\sum_{i\supset v}
\alpha_i =2\pi -\delta$, for $\delta\not= 0$, see Fig.\ref{euangle}.
The so-called deficit angle $\delta$ 
is precisely the rotation angle picked up by the vector and is
a direct measure for the scalar curvature at the vertex. The
operational description to obtain the scalar curvature in higher
dimensions is very similar, one basically has to sum in each point
over the Gaussian curvatures of all two-dimensional submanifolds.
This explains why in Regge calculus the curvature part of the
Einstein action is given by a sum over building blocks of
dimension $(d-2)$ which are simply the objects dual to
those local 2d submanifolds. More precisely, the continuum
curvature and volume terms of the action become
\begin{eqnarray}
\frac{1}{2} \int_{\cal R} d^dx\ \sqrt{ |\det g |} {}^{(d)}R &\longrightarrow&
\sum_{i\in {\cal R}} Vol(i^{th}\ (d-2){\rm -simplex})\ \delta_i \label{curv}\\
 \int_{\cal R} d^dx\ \sqrt{ |\det g |} &\longrightarrow&
\sum_{i\in {\cal R}} Vol(i^{th}\ d{\rm -simplex})\label{vol}
\end{eqnarray}
in the simplicial discretization. It is then a simple exercise in
trigonometry to express the volumes and angles appearing in
these formulas as functions of the edge lengths $l_i$, both in
the Euclidean and the Minkowskian case. 

The approach of dynamical triangulations uses a certain class of
such simplicial space-times as an explicit, regularized realization of the
space Geom($M$). For a given volume $N_d$, this class consists 
of all gluings of manifold-type of a set of $N_d$ simplicial building 
blocks of top-dimension $d$ whose edge lengths are restricted to take
either one or one out of two values. In the Euclidean
case we set $l_i^2=a^2$ for all $i$, and in the Lorentzian case
we allow for both space- and time-like links with $l_i^2\in\{-a^2,a^2\}$,
where the geodesic distance $a$ serves as a short-distance
cutoff, which will be taken to zero later. Coming from the classical
theory this may seem a grave restriction at first, but this is indeed
not the case. Firstly, keep in mind that for the purposes of the
quantum theory we want to sample the space of geometries ``ergodically''
at a coarse-grained scale of order $a$. This should be contrasted
with the classical theory where the objective is usually to
approximate a given, {\it fixed} space-time to within a length
scale $a$. In the latter case one typically requires a much
finer topology on the space of metrics or geometries. It is
also straightforward to see that no local curvature degrees of
freedom are suppressed by fixing the edge lengths; deficit
angles in all directions are still present, although they 
take on only a discretized set of values. In this sense, in
dynamical triangulations all geometry is in the gluing of the
fundamental building blocks. This is dual to how
quantum Regge calculus is set up, where one usually fixes a 
triangulation $T$ and then ``scans'' the space of geometries by
letting the $l_i$'s run continuously over all values  
compatible with the triangular inequalities.

In a nutshell, Lorentzian dynamical triangulations give a
definite meaning to the ``integral over geometries'', namely,
as a sum over inequivalent Lorentzian gluings $T$ over any number
$N_d$ of $d$-simplices,
\begin{equation}
\int_{\rm Geom(M)} {\cal D}[g_{\mu\nu}]\ 
{\rm e}^{i S [g_{\mu\nu}]}\;\; \;\;\; \stackrel {\rm LDT}{\longrightarrow}
\;\;\;\;\;
\sum_{T \in {\cal T}}\frac{1}{C_T}\  {\rm e}^{i S^{\rm Regge}(T)},
\label{nut}
\end{equation}
where the symmetry factor $C_T=|Aut(T)|$ on the right-hand side 
is the order of the automorphism group of the triangulation,
consisting of all maps of $T$ onto
itself which preserve the connectivity of the simplicial lattice.
I will specify below what precise class $\cal T$ of triangulations
should appear in the summation.

It follows from the above that in this formulation all curvatures and volumes
contributing to the simplicial Regge action come in discrete
units. This is again easily illustrated by the case of a
two-dimensional triangulation with Euclidean signature,
which according to the prescription of dynamical triangulations
consists of {\it equilateral} triangles with squared edge
lengths $+a^2$. All interior angles of such a triangle are equal
to $\pi/3$, which implies that the deficit angle at any vertex
$v$ can take the values $2\pi-k_v \pi/3$, where $k_v$ is the
number of triangles meeting at $v$. As a consequence, the
Einstein-Regge action assumes the simple form\footnote{Strictly speaking,
the expression (\ref{dtaction}) in $d\geq 3$ is only correct for
the Euclidean or the Wick-rotated Lorentzian action. In the 
Lorentzian case one has several types of simplices of a given
dimension $d$, depending on how many of its links are time-like.
Only after the Wick rotation will all links be space-like and of equal length
(see later). Nevertheless, I will use this more compact form for
ease of notation.}
\begin{equation}
S^{\rm Regge}(T)=\kappa_{d-2} N_{d-2}-\kappa_d N_d,
\label{dtaction}
\end{equation}
where the coupling constants $\kappa_i=\kappa_i (\lambda,
G_N)$ are simple functions of the bare cosmological
and Newton constants in $d$ dimensions. Substituting
this into the path sum in (\ref{nut}) leads to
\begin{equation}
Z(\kappa_{d-2},\kappa_d)=\sum_{N_d} {\rm e}^{-i\kappa_d N_d}
\sum_{N_{d-2}}{\rm e}^{i\kappa_{d-2} N_{d-2}}\sum_{T|_{N_d,
N_{d-2}}} \frac{1}{C_T},
\label{combact}
\end{equation}
The point of
taking separate sums over the numbers of $d$- and $(d-2)$-simplices
in (\ref{combact}) is to make explicit that ``doing the sum'' is tantamount
to the combinatorial problem of {\it counting} triangulations of a given
volume and number of simplices of co-dimension two (corresponding
to the last summation in (\ref{combact})).\footnote{The symmetry factor 
$C_T$ is
almost always equal to 1 for large triangulations.} It turns out that at least
in two space-time dimensions the counting of geometries can be
done completely explicitly, turning both Lorentzian and Euclidean
quantum gravity into exactly soluble statistical models. 

\begin{figure}[h]
\vspace{0.4cm}
\centerline{\scalebox{0.65}{\rotatebox{0}
{\includegraphics{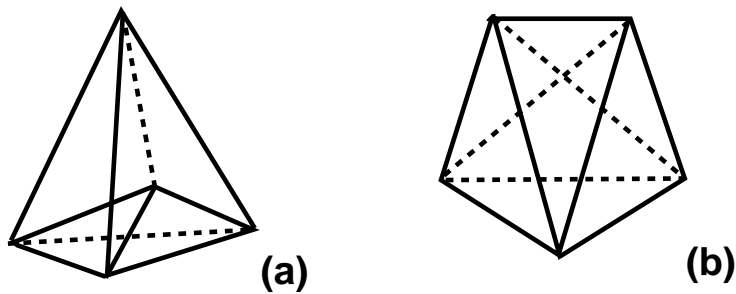}}}}
\caption[4dsimplex]{The two types of Minkowskian four-simplices in four
dimensions.}
\vspace{0.5cm}
\label{4dsimplex}
\end{figure}

\section{Lorentzian nature of the path integral}\label{lorentz}

It is now time to explain what makes our approach {\it Lorentzian}
and why it therefore differs from previous attempts at
constructing non-perturbative gravitational path integrals.
The simplicial building blocks of the models are taken to be
pieces of Minkowski space, and their edges have squared lengths
$+a^2$ or $-a^2$. For example, the two types of four-simplices  
that are used in Lorentzian dynamical triangulations in dimension
four are shown in Fig.\ref{4dsimplex}. 
The first of them has four time-like and
six space-like links (and therefore contains 4 time-like and 1
space-like tetrahedron), whereas the second one has six time-like
and four space-like links (and contains 5 time-like
tetrahedra). Since both are subspaces of flat space
with signature $(-+++)$, they possess well-defined light-cone
structures everywhere. 

In general, gluings between pairs of 
$d$-simplices are only possible when the metric
properties of their $(d-1)$-faces match. Having local light cones
implies causal relations between pairs of points in local
neighbourhoods. Creating closed time-like
curves will be avoided by requiring that all space-times contributing to the
path sum possess a global ``time'' function $t$. In terms of the
triangulation this means that the $d$-simplices are 
arranged such that their space-like links all lie in slices
of constant integer $t$, and their time-like links
interpolate between adjacent spatial slices $t$ and $t+1$.
Moreover, with respect to this time, we will not allow for
any {\it spatial} topology changes\footnote{Note that if we were
in the continuum and had introduced coordinates on space-time,
such a statement would actually be diffeomorphism-invariant.}.

\begin{figure}[h]
\vspace{0.5cm}
\centerline{\scalebox{0.4}{\rotatebox{0}
{\includegraphics{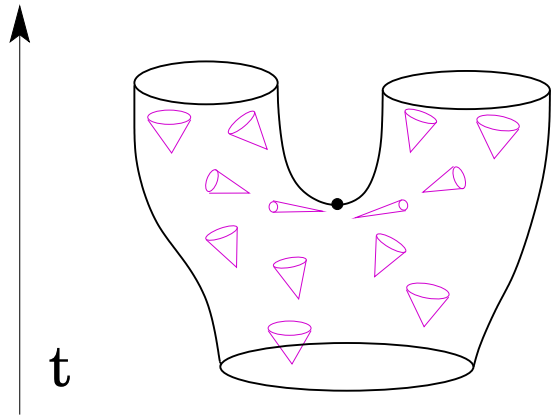}}}}
\vspace{0.5cm}
\caption[trousers]{At a branching point associated with
a spatial topology change, light-cones get ``squeezed''.}
\label{trousers}
\end{figure} 

This latter condition is always satisfied in classical
applications, where ``trouser points'' like the one
depicted in Fig.\ref{trousers} are ruled
out by the requirement of having a non-degenerate 
Lorentzian metric defined everywhere on $M$ (it is geometrically
obvious that the light cone and hence $g_{\mu\nu}$ 
must degenerate in at least
one point along the ``crotch''). Another way of thinking about
such configurations (and their time-reversed counterparts)
is that the causal past (future) of an observer changes
discontinuously as her worldline passes near the singular point
(see \cite{dowker} and references therein for related discussions 
about the issue of topology change in quantum gravity). 

Of course, there is no
{\it a priori} reason in the quantum theory to not relax
some of these classical causality constraints. After all,
as I stressed right at the outset, path
integral histories are not in general classical solutions,
nor can we attribute any other direct physical meaning to
them individually. It might well be that one can
construct models whose path integral configurations
violate causality in this strict sense, but where this notion
is somehow recovered in the resulting continuum theory.
What the approach of Lorentzian dynamical triangulations
has demonstrated is that {\it imposing causality constraints
will in general lead to a different continuum theory}.
This is in contrast with the intuition one may have
that ``including a few isolated singular points will not
make any difference''. On the contrary, tampering with
causality in this way is not innocent at all, as was already anticipated
by Teitelboim many years ago \cite{teitelboim}. 

I want to point out that one cannot conclude from the above that 
spatial topology changes or even fluctuations in the {\it space-time}
topology cannot be treated in the formulation of
dynamical triangulations. However, if one insists on including
geometries of variable topology in a Lorentzian discrete
context, one has to come up with a prescription of how to
weigh these singular points in the path integral, both before and 
after the Wick 
rotation. Maybe this can be done along the lines suggested in
\cite{loukosorkin}; this is clearly an interesting issue for further research.

Having said this, we next have to address the question of the
Wick rotation, in other words, of how to get rid of the factor of
$i$ in the exponent of (\ref{combact}). Without it, this expression
is an infinite sum (since the volume can become arbitrarily
large) of complex terms whose convergence properties will
be very difficult to establish. In this situation, a Wick rotation is 
simply a technical tool which -- in the best of all worlds -- 
enables us to perform the state sum and determine its
continuum limit. Of course, the end result will have to be
Wick-rotated back to Lorentzian signature. 

Fortunately, Lorentzian dynamical triangulations come with
a natural notion of Wick rotation, and the strategy I just
outlined can be carried out explicitly in two space-time
dimensions, leading to a unitary theory (see Sec.\ \ref{two} below).
In higher dimensions we do not yet have sufficient analytical
control of the continuum theories to make specific statements 
about the {\it inverse} Wick
rotation. Since we use the Wick rotation at an intermediate
step, one can ask whether other Wick rotations would
lead to the same result. Currently this is a somewhat academic
question, since it is in practice difficult to find such alternatives. 
In fact, it is quite miraculous we have found a single prescription
for Wick-rotating in our regularized setting, and it does not seem
to have a direct continuum analogue  
(for more comments on this issue, see \cite{conformal,dasgupta}). 

Our Wick rotation $W$ in any dimension is an injective map from 
Lorentzian- to Euclidean-signature simplicial space-times. 
Using the notation {\tt T} for a simplicial manifold together
with length assignments $l_s^2$ and $l_t^2$ to its space- 
and time-like links, it is defined by 
\begin{equation}
{\tt T}^{\rm lor}=(T,\{ l_s^2=a^2, l_t^2=-a^2\} )\;\;\stackrel{W}{\mapsto}
\;\; {\tt T}^{\rm eu}=(T,\{ l_s^2=a^2, l_t^2= a^2\} ).
\label{wick}
\end{equation}
Note that we have not touched the connectivity of the simplicial
manifold $T$, but only its metric properties, by mapping 
all time-like links of $T$ into
space-like ones, resulting in a Euclidean ``space-time" of 
equilateral building blocks.
It can be shown \cite{d3d4} that at the level of the corresponding weight 
factors in the path integral this Wick rotation\footnote{To obtain a genuine
Wick rotation and not just a discrete map, one introduces a complex
parameter $\alpha$ in $l_t^2 =-\alpha a^2$. The proper prescription
leading to (\ref{wickact}) is then an analytic continuation of $\alpha$
from 1 to $-1$ through the lower-half complex plane.} has precisely the
desired effect of rotating to the exponentiated Regge action of the
Euclideanized geometry,
\begin{equation}
{\rm e}^{iS({\cal T}^{\rm lor})}\;\; \stackrel{W}{\mapsto}\;\;
{\rm e}^{-S({\cal T}^{\rm eu})}.
\label{wickact}
\end{equation}

The Euclideanized path sum after the Wick rotation has the form
\begin{eqnarray}
Z^{\rm eu}(\kappa_{d-2},\kappa_d)&=&
\sum_T \frac{1}{C_T}\ {\rm e}^{-\kappa_d N_d (T) +\kappa_{d-2} N_{d-2}(T)}
\nonumber \\
&= &\sum_{N_d} {\rm e}^{-\kappa_d N_d}
\sum_{T|_{N_d}} \frac{1}{C_T}\ {\rm e}^{\kappa_{d-2} N_{d-2}(T)} \nonumber\\
&=& \sum_{N_d} {\rm e}^{-\kappa_d N_d}\ 
{\rm e}^{\kappa_d^{\rm crit}(\kappa_{d-2})N_d}\times 
{\rm subleading}(N_d).
\label{euclact}
\end{eqnarray}
In the last equality I have used that the number of Lorentzian
triangulations of discrete volume $N_d$ to leading order scales
exponentially with $N_d$ for large volumes. This can be shown
explicitly in space-time dimension 2 and 3. For $d=4$, there
is strong (numerical) evidence for such an exponential bound for
{\it Euclidean} triangulations, from which the desired result
for the Lorentzian case follows (since $W$ maps to a 
strict subset of all Euclidean simplicial manifolds).

From the functional form of the last line of (\ref{euclact}) one
can immediately read off some qualitative features of the phase
diagram, an example of which appeared already earlier in
Fig.\ref{phaseeu}. Namely, the sum over geometries $Z^{\rm eu}$
converges for values $\kappa_d >\kappa_d^{\rm crit}$ of the
bare cosmological constant, and diverges (ie. is not defined)
below this critical line. Generically, for all models of
dynamical triangulations the infinite-volume limit 
is attained
by approaching the critical line $\kappa_d^{\rm crit}(\kappa_{d-2})$
from above, ie. from inside the region of convergence of
$Z^{\rm eu}$. In the process of taking $N_d\rightarrow\infty$ and
the cutoff $a\rightarrow 0$, one obtains a renormalized
cosmological constant $\Lambda$ through
\begin{equation}
(\kappa_d -\kappa_d^{\rm crit})=a^\mu\Lambda +O(a^{\mu +1}).
\label{lambdas}
\end{equation}
If the scaling is canonical (which means that the dimensionality
of the renormalized coupling constant is the one expected from
the classical theory), the exponent is given by $\mu=d$. 
Note that this construction
requires a positive {\it bare} cosmological constant in order to
make the state sum converge. Moreover, by virtue of relation 
(\ref{lambdas})
also the {\it renormalized} cosmological constant must be positive. 
Other than that, its
numerical value is not determined by this argument, but
by comparing observables of the theory which depend on $\Lambda$
with actual physical measurements.\footnote{The non-negativity of the 
renormalized
cosmological coupling may be taken as a first ``prediction" of our
construction, which in the physical case of four dimensions is indeed in 
agreement with current observations.} Another interesting observation 
is that the inclusion of a sum over topologies in the discretized sum 
(\ref{euclact}) would lead to a super-exponential growth of at least $\propto N_d!$ 
of the number of triangulations with the volume $N_d$. Such a divergence
of the path integral cannot be compensated by an additive renormalization
of the cosmological constant of the kind outlined above. 

There are of course ways in which one can sum divergent series of this 
type, for example, by performing a Borel sum. The problem with 
these stems from the fact that two different functions can share the
same asymptotic expansion. Therefore, the series in itself is {\it not}
sufficient to define the underlying theory uniquely. The non-uniqueness
arises because of non-perturbative contributions to the path integral
which are not represented in the perturbative expansion.\footnote{A
field-theoretic example would be instantons and renormalons in QCD.} 
In order to fix
these uniquely, an independent, non-perturbative definition of the
theory is necessary. Unfortunately, for dynamically
triangulated models of quantum gravity, no such definitions have
been found so far. In the context of
two-dimensional (Euclidean) quantum gravity this difficulty is 
known as the ``absence of a physically motivated double-scaling
limit" \cite{doublescaling}. The same issue has recently been revived
in $d=3$ \cite{freidel}, where the situation is not any better.

Lastly, obtaining an interesting continuum limit may
or may not require an additional fine-tuning of the inverse
gravitational coupling $\kappa_{d-2}$, depending on the dimension
$d$. In four dimensions, one would expect to find
a second-order transition along the critical line, corresponding to
local gravitonic excitations. The situation in $d=3$ is less clear,
but results obtained so far indicate that no fine-tuning of 
Newton's constant is necessary \cite{ajl2,ajl3}. 

Before delving into the details, let me summarize briefly the
results that have been obtained so far in the approach of
Lorentzian dynamical triangulations. At the regularized level,
that is, in the presence of a finite cutoff $a$ for the edge lengths and
an infrared cutoff for large space-time volume, they are well-defined
statistical models of Lorentzian random geometries in
$d=2,3,4$. In particular, they obey a suitable notion of
reflection-positivity and possess selfadjoint Hamiltonians. 

The crucial questions are then to what extent the underlying
combinatorial problems of counting all $d$-dimensional
geometries with certain causal properties 
can be solved, whether continuum
theories with non-trivial dynamics exist and how their bare
coupling constants get renormalized in the process.
What we know about Lorentzian dynamical triangulations so far
is that they lead to continuum theories
of quantum gravity in dimension 2 and 3. 
In $d=2$, there is a complete analytic
solution, which is distinct from the continuum theory
produced by Euclidean dynamical triangulations. Also
the matter-coupled model has been studied. In $d=3$, there are
numerical and partial analytical results which show that both a
continuum theory exists and that it again differs from its
Euclidean counterpart. Work on a more complete analytic solution
which would give details about the geometric properties of the
quantum theory is under way. In $d=4$, the first numerical
simulations are currently being set up. The challenge here is
to do this for sufficiently large lattices, to be able to perform
meaningful measurements. So far, we cannot make any
statements about the existence and properties of a continuum
theory in this physically most interesting case.

\subsection{In two dimensions}\label{two}

The two-dimensional case serves as a nice illustration of the 
objectives of the approach, many of which can be carried out in a
completely explicit manner \cite{al}.
There is just one type of building block, a flat Minkowskian
triangle with two time-like edges of squared edge lengths
$l_t^2=-a^2$ and one space-like edge with $l_s^2=a^2$.
We
build up a causal space-time from strips of unit height
$\Delta t=1$ (see Fig.\ref{2dtriang}), 
where $t$ is an integer-valued discrete parameter
that labels subsequent spatial slices, ie. simplicial
submanifolds of codimension 1 which are constructed from
space-like links only. In the two-dimensional case these subspaces
are one-dimensional. We choose periodic boundary conditions,
such that the spatial ``universes'' are topologically spheres $S^1$
(other boundary conditions are also possible, leading to a
slight modification of the effective quantum Hamiltonian
\cite{krist1,krist2}). A spatial geometry at given $t$ is completely
characterized by its length $l(t)\in\{1,2,3,\dots\}$, which (in units of the
lattice spacing $a$) is simply the number of spatial
edges it contains.

\begin{figure}[h]
\vspace{0.5cm}
\centerline{\scalebox{0.5}{\rotatebox{0}
{\includegraphics{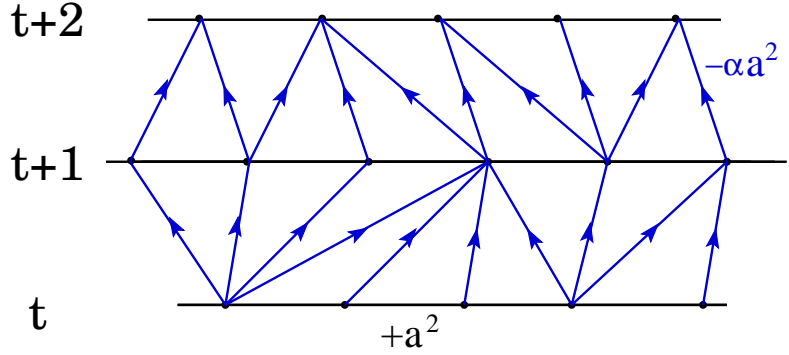}}}}
\vspace{0.5cm}
\caption[2dtriang]{Two strips of a 2d Lorentzian triangulation,
with spatial slices of constant $t$ 
and interpolating future-oriented time-like
links.}
\label{2dtriang}
\end{figure} 

One simplification occurring in two dimensions is that the
curvature term in the Einstein action is a topological invariant
(and that therefore does not depend on the metric), given by
\begin{equation}
\int_M d^2x\ \sqrt{|\det g|}R=2\pi\chi,
\label{rtwo}
\end{equation}
where $\chi$ denotes the Euler characteristic of the two-dimensional
space-time $M$. Since we are keeping the space-time topology fixed,
the exponential of $i$ times this term is a constant overall
factor that can be pulled out of the
path integral and does not contribute to the dynamics. Dropping this
term, we can write the discrete path integral over 2d simplicial
causal space-times as
\begin{equation}
G_\lambda(l_{\rm in},l_{\rm out};t)=\sum_{ {{\rm causal}\,\, T}\atop
{l_{\rm in},l_{\rm out},t}} {\rm e}^{-i\lambda N_2}
\;\;\stackrel{\rm Wick}{\longrightarrow} \;\;
\sum_{ {W( T)}\atop
{l_{\rm in},l_{\rm out},t}} {\rm e}^{-\tilde\lambda N_2},
\label{prop2d}
\end{equation}
where the weight factors depend now only on the cosmological
(volume) term, and $\tilde\lambda$ differs from $\lambda$ by
a finite positive numerical factor. Each history entering in the discrete
propagator (\ref{prop2d}) has an in-geometry of length $l_{\rm in}$,
an out-geometry of length $l_{\rm out}$, and 
consists of $t$ steps. An important special
case is the propagator for a single step, which in its Wick-rotated
form reads\footnote{This is the ``unmarked'' propagator, 
see \cite{al,iceland} for details.}
\begin{equation}
G_{\tilde\lambda}(l_1,l_2;t=1)=\langle l_2|\hat T|l_1
\rangle ={\rm e}^{-\tilde\lambda (l_1+l_2)} \sum_{T:l_1
\rightarrow l_2} 1\equiv {\rm e}^{-\tilde\lambda (l_1+l_2)}
\frac{1}{l_1+l_2}\Biggl( {l_1 +l_2  \atop l_1}\Biggr).
\label{prop1s}
\end{equation}
The second equation in (\ref{prop1s}) defines the transfer matrix
$\hat T$ via its matrix elements in the basis of the (improper)
length eigenvectors $|l\rangle$. Knowing the eigenvalues of the
transfer matrix is tantamount to a solution of the general
problem by virtue of the relation
\begin{equation}
G_{\tilde\lambda}(l_1,l_2;t)=\langle l_2|\hat T^t|l_1
\rangle.
\label{propmanys}
\end{equation}
Importantly, the propagator satisfies the composition property
\begin{equation}
G_{\tilde\lambda}(l_1,l_2;t_1+t_2)=
\sum_{l=1}^\infty G_{\tilde\lambda}(l_1,l;t_1)l
G_{\tilde\lambda}(l,l_2;t_2),
\label{compos}
\end{equation}
where the sum on the right-hand side is over a complete set of
intermediate length eigenstates.

\begin{figure}[h]
\vspace{0.5cm}
\centerline{\scalebox{0.6}{\rotatebox{0}
{\includegraphics{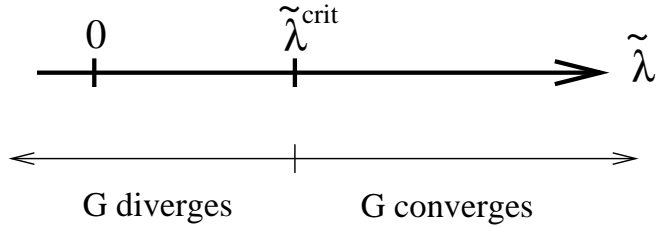}}}}
\vspace{0.5cm}
\caption[2dphase]{The 1d phase diagram of 2d Lorentzian
dynamical triangulations.}
\label{2dphase}
\end{figure} 

Next, we look for critical behaviour of the propagator
$G_{\tilde\lambda}$ (that is, a non-analytic behaviour as a
function of the renormalized coupling constant) in the limit as
$a\rightarrow 0$. Since there is only one coupling,
the phase diagram of the theory is just one-dimensional, and
illustrated in Fig.\ref{2dphase}. As can be read off from 
the explicit form of the propagator,
\begin{equation}
G_{\tilde\lambda}=\sum_{N_2} {\rm e}^{-\tilde\lambda N_2}
\sum_{T|_{N_2}} 1=\sum_{N_2} {\rm e}^{-(\tilde\lambda-
\tilde\lambda^{\rm crit}) N_2}\times {\rm subleading}(N_2),
\label{prop2dagain}
\end{equation}
the discrete sum over 2d geometries converges above some
critical value $\tilde\lambda^{\rm crit}>0$, and diverges
for $\tilde\lambda$ below this point. In order to attain a
macroscopic physical volume $\langle V\rangle :=\langle
a^2N_2 \rangle$ in the $a\rightarrow 0$ limit, one needs to
approach $\tilde\lambda^{\rm crit}$ from above. It turns out
that to get a non-trivial continuum limit, the bare
cosmological coupling constant has to be fine-tuned
canonically according to
\begin{equation}
\tilde\lambda -\tilde\lambda^{\rm crit}=a^2 \Lambda^{\rm ren}+
O(a^3).
\label{lrenorm}
\end{equation}
Note that the numerical value of $\tilde\lambda^{\rm crit}$ 
will depend on the details of the discretization (for example,
the building blocks chosen; see \cite{krist1} for alternative choices),
the so-called non-universal properties of the model which do
not affect the quantum dynamics of the final continuum theory.
At the same time, the counting variables $l$ and $t$ are
taken to infinity while keeping the dimensionful quantities
$L:=al$ and $T:=at$ constant. The renormalized propagator
is then defined as a function of all the renormalized
variables,
\begin{equation}
G_\Lambda (L_1,L_2;T):= \lim_{a\rightarrow 0} a^\nu
G_{\tilde\lambda^{\rm crit}+a^2\Lambda}\Bigl(\frac{L_1}{a},
\frac{L_2}{a};\frac{T}{a}\Bigr),
\label{fullprop}
\end{equation}
which also contains a multiplicative wave function
renormalization. The final result for the continuum path
integral of two-dimensional Lorentzian quantum gravity is
obtained by an inverse Wick rotation of the continuum
proper time $T$ to $iT$ from the Euclidean expression and is given by
\begin{equation}
G_\Lambda (L_{\rm in},L_{\rm out};T)={\rm e}^{-\coth (i
\sqrt{\Lambda} T)\sqrt{\Lambda}( L_{\rm in}+L_{\rm out})}\,
\frac{\sqrt{\Lambda L_{\rm in}L_{\rm out}}}{\sinh
(i\sqrt{\Lambda} T)}\;
I_1\Biggl(
\frac{2\sqrt{\Lambda L_{\rm in}L_{\rm out}}}{\sinh
(i\sqrt{\Lambda} T)}\Biggr),
\label{finalprop}
\end{equation}
where $I_1$ denotes the Bessel function of the first kind.

\begin{figure}[h]
\vspace{0.5cm}
\centerline{\scalebox{0.8}{\rotatebox{0}
{\includegraphics{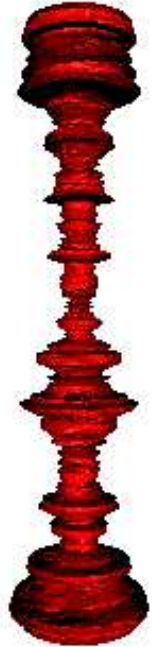}}}}
\vspace{0.5cm}
\caption[puregrav]{A typical two-dimensional Lorentzian space-time, with volume
$N_2=18816$ and a total proper time of $t=168$ steps.}
\label{puregrav}
\end{figure}

What is the physics behind this functional expression? In two dimensions,
there is not much ``physics" in the sense that
the {\it classical} Einstein equations are empty. This renders
meaningless 
the question of a classical limit of the 2d quantum theory;
whatever dynamics there is will be purely ``quantum". Fig.\ref{puregrav}
shows a typical two-dimensional quantum universe: 
the compactified direction
is ``space", and the vertical axis is ``time". It illustrates the
typical development of the ground state of the system over
time, as generated by a Monte-Carlo simulation of almost 
19.000 triangles. 

Since the theory has been solved analytically, we also 
know the explicit
form of the effective quantum Hamiltonian, namely,
\begin{equation}
\hat H=-L\frac{d^2}{dL^2}-2\frac{d}{dL}+\Lambda L.
\label{2dham}
\end{equation}
This operator is selfadjoint on the Hilbert space
$L^2({\bf R}_+,LdL)$ and generates a unitary evolution in
the continuum proper time $T$. The Hamiltonian consists of
a kinetic term in the single geometric variable $L$
(the size of the spatial universe) and a potential
term depending on the renormalized cosmological
constant. Its spectrum is discrete,
\begin{equation}
E_n=2(n+1)\sqrt{\Lambda},\;\;\; n=0,1,2,\dots
\label{spectrum}
\end{equation}
and one can compute various expectation values, for
example,
\begin{equation}
\langle L\rangle_n=\frac{n+1}{\sqrt{\Lambda}},
\;\;\;\; \;\;
\langle L^2 \rangle_n=\frac{3}{2}\frac{(n+1)^2}{\Lambda}.
\label{expect}
\end{equation}
Since there is just one dimensionful constant, with $[\Lambda]=
{\rm length}^{-2}$, all dimensionful
quantities must appear in appropriate units of $\Lambda$.

Another useful way of characterizing the continuum theory
is via certain critical exponents, which in the case of gravitational
theories are of a geometrical nature. The {\it Hausdorff dimension}
$d_H$ describes the scaling of the volume of a geodesic ball of
radius $R$ as a function of $R$. This very general notion can be
applied to a fixed metric space, but for our purposes we are 
interested in the ensemble average over the entire ``sum over
geometries", that is, the leading-order scaling behaviour of the expectation 
value\footnote{For the Lorentzian theory, ``geodesic distance"
refers to the length measurements after the Wick rotation.} 
\begin{equation}
\langle V(R)\rangle \propto R^{d_H}.
\label{hausdorff}
\end{equation}

The Hausdorff dimension is a truly dynamical quantity, and is
{\it not} a priori the same as the dimensionality of the building
blocks that were used to construct the individual discrete
space-times in the first place. It may even depend on the
length scale of the radial distance $R$. Remarkably,
$d_H$ can be calculated analytically in both Lorentzian and
Euclidean 2d quantum gravity (see, for example, \cite{alnr}). 
The latter, also known as
``Liouville gravity", can be obtained by performing a sum
over {\it arbitrary} triangulated Euclidean two-geometries 
(with fixed topology $S^2$), and not just those which correspond to
a Wick-rotated causal Lorentzian space-time. 
One finds
\begin{equation}
d_H=2\;\;\; ({\rm Lorentzian})\;\;\;\;\;\; {\rm and} \;\;\;\;\;\;
d_H=4\;\;\; ({\rm Euclidean}).
\label{dims}
\end{equation}
The geometric picture associated with the non-canonical
value of $d_H$ in the Euclidean case is that of a fractal geometry, 
with wildly branching ``baby universes". This branching
behaviour is incompatible with the causal structure required in
the Lorentzian case, and the geometry of the Lorentzian 
quantum ground state
is much better behaved, although it is by no means smooth as
we have already seen. 

We conclude that the continuum theories of 2d quantum gravity
with Euclidean and Lorentzian signature are distinct. They can
be related by a somewhat complicated renormalization procedure
which one may think of as ``integrating out the baby universes"
\cite{ackl}, which is not at all as simple as ``sticking a factor of $i$ in the
right place". In a way, this is not unexpected in view of the fact
that (the spaces of) Euclidean and Lorentzian geometries are
already classically very different objects. I am not claiming
that from the point of view of 2d quantum gravity, one signature 
is better than the other. This seems a matter of taste, since neither 
theory describes any aspects of real nature. 
Nevertheless, what we have shown is that
imposing causality constraints at the level of the individual histories
in the path integral changes the outcome radically, a feature one
may expect to generalize to higher dimensions.  

Let me comment at this point about the role of the integer $t$
which labels the time steps in the propagator (\ref{propmanys})
and its higher-dimensional analogues. In the first place, it is
one of the many discrete parameters that label the regularized
space-times {\it in a coordinate-invariant way}. In any given
Minkowskian building block, one may introduce proper-time
coordinates whose value coincides (up to a constant factor
depending on the type of the building block) with the discrete
time $t$ on the spatial slices. However, this is where the analogue
with continuum proper time ends, since it is in general impossible
to extend such coordinate patches over more than one time step,
because of the presence of curvature singularities. Next, there is
no claim that the propagator with respect to $t$ or its continuum 
analogue $T$ has a distinguished
{\it physical} meaning, despite being invariantly defined. 
Nevertheless, we do believe strongly that it contains all
physical information about the ``quantum geometry". In other
words, all observables and propagators (which may depend
on other notions of ``time") can in principle be computed 
from our propagator in $t$.\footnote{A related result
has already been demonstrated for the proper-time propagator in 
two-dimensional {\it Euclidean} quantum gravity \cite{kawai}.}
This can of course be difficult in practice, but this is only to be
expected.

\vspace{.8cm}

Coming from Euclidean quantum gravity, there are specific reasons
for looking at the behaviour of the
matter-coupled theory in two dimensions. 
The coupling of matter fields to Lorentzian
dynamical triangulations can be achieved in the usual manner by
including for each given geometry $T$ in the path integral a 
summation over all matter degrees of freedom on $T$,
resulting in a double sum over geometric and matter variables.
For example, adding Ising spins to 2d Lorentzian gravity is
described by the partition function
\begin{equation}
Z(\lambda,\beta_I)=\sum_{N_2} {\rm e}^{-\lambda N_2}
\sum_{{\rm causal}\atop T\in {\cal T}_{N_2}} 
\sum_{\{\sigma_i=\pm 1\} }{\rm e}^{\frac{\beta_I}{2} \sum_{<ij>} 
\sigma_i\sigma_j},
\label{ising}
\end{equation}
where the last sum on the right is over the spin configurations 
of the Ising model on the
triangulation $T$. The analogous model on Euclidean triangulations
has been solved exactly \cite{boulatov}, 
and its continuum matter behaviour is
characterized by the critical exponents
\begin{equation}
\alpha=-1,\;\;\;\; \beta=0.5,\;\;\;\; \gamma=2,\;\;\;\;\;\;
({\rm Euclidean})
\label{eucrit}
\end{equation}
for the specific heat, the magnetization and the
magnetic susceptibility respectively. These differ from the ones 
found for the Ising model on a fixed, flat lattice, 
the so-called Onsager exponents. The transition here is
third-order, reflecting the influence of the fractal background
on which the matter is propagating. 

The same Ising model, when coupled to Lorentzian geometries
according to (\ref{ising}), has not so far been solved exactly,
but its critical matter exponents have been determined numerically
and by means of a diagrammatic high-$T$ expansion \cite{aal1} and 
agree (within error bars) with the Onsager exponents, that is,
\begin{equation}
\alpha=0,\;\;\;\; \beta=0.125,\;\;\;\; \gamma=1.75.\;\;\;\;\;\;
({\rm Lorentzian})
\label{lorcrit}
\end{equation}
So, interestingly, despite the fluctuations of the geometric
ensemble evident in Fig.\ref{puregrav}, the conformal matter behaves
{\it as if} it lived on a static flat lattice. This indicates a certain
robustness of the Onsager behaviour in the presence of such 
fluctuations. Does it also imply there cannot be any 
back-reaction of the matter on the geometry? In order to
answer this question, Lorentzian quantum gravity was coupled
to ``a lot of matter", in this case, eight copies of Ising models \cite{aal2}.
The partition function is a direct generalization of (\ref{ising}).
For a given triangulation, there are 8 independent Ising models,
which interact with each other only via their common
interaction with the ensemble of geometries.                      
\begin{figure}[h]
\vspace{0.5cm}
\centerline{\scalebox{0.6}{\rotatebox{0}
{\includegraphics{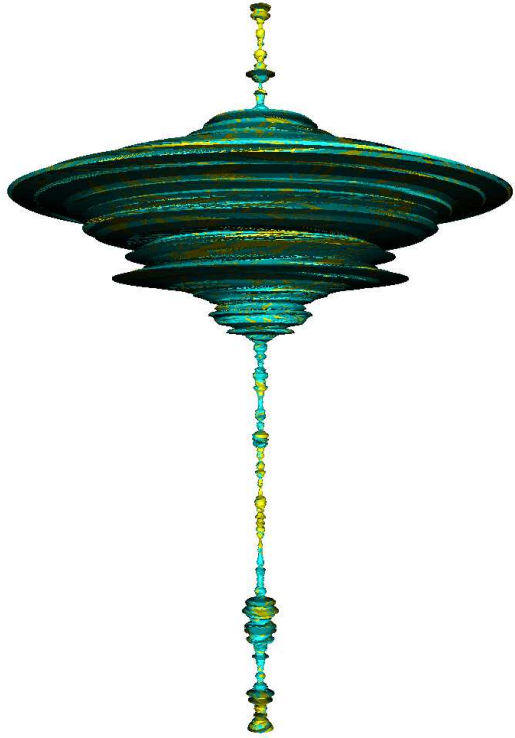}}}}
\vspace{0.5cm}
\caption[ising8]{A typical two-dimensional Lorentzian geometry 
in the presence of eight
Ising models, for volume $N_2=73926$ and a total proper time $t=333$.}
\label{ising8}
\end{figure}

Looking again at a typical ``universe'', depicted in Fig.\ref{ising8}, 
its geometry is now significantly changed in comparison with the case without
matter. Part of it is squeezed down to a spatial universe of
minimal size, with the remainder forming a genuinely
extended space-time. A measurement of the critical behaviour
of the matter on this piece of the universe again produces
values compatible with the Onsager exponents!\footnote{The same would of
course not hold for the degenerate part of the space-time which is
effectively one-dimensional.}
This is a very interesting result from the point of view of Liouville
gravity, which does not seem to produce meaningful matter-coupled
models beyond a central charge of one, the famous $c=1$ barrier.
(A model with $n$ Ising spins corresponds to central charge $c=n/2$.)
We conclude that  {\it causal} space-times are better carrier
spaces for matter fields in 2d quantum gravity.

\subsection{In three dimensions}\label{three}

Having discovered the many beautiful features of being Lorentzian
in two dimensions, the next challenge is to solve the dynamically
triangulated model in three dimensions and understand the
geometric properties of the continuum theory it gives rise to.
This will bring us a step closer to our ultimate goal, the
four-dimensional quantum theory. 

Despite its reputation as an ``exactly soluble theory", many
aspects of quantum gravity in 2+1 dimensions remain to be understood.
There is still an unresolved tension between (i) the gauge (Chern-Simons) 
formulation in which the constraints can be solved in a
straightforward way before or after quantization, leading to a
quantized finite-dimensional phase space, and (ii) a path integral
formulation in terms of ``$g_{\mu\nu}$" which seems just about as
intractable as the four-dimensional theory, and is power-counting
non-renormalizable. 

Since Lorentzian dynamical triangulations are really a regularized
and non-per\-tur\-ba\-tive version of the latter, a solution of the model
should help to bridge this gap. Part of the trouble with gravitational
path integrals is the ``conformal-factor problem", which makes its
first appearance in $d=3$.\footnote{A more detailed account of the
history of this problem in quantum gravity can be found in \cite{conformal}.}
The conformal part of the metric, ie.
the mode associated with an overall scaling of all components of
the metric tensor, contributes to the action with a kinetic term of
the wrong sign. This is most easily seen by considering just the
curvature term of the Einstein action,
\begin{equation}
S=\int d^dx \sqrt{g} (R+\dots),
\label{conf1}
\end{equation}
and performing a conformal transformation $g_{\mu\nu}\rightarrow
g_{\mu\nu}'={\rm e}^{\phi}g_{\mu\nu}$ on the metric. This is {\it not}
a gauge transformation and leads to a change 
\begin{equation}
S\rightarrow S'=\int d^dx \sqrt{g'} (-(\partial_0\phi)^2 +\dots)
\label{conf2}
\end{equation}
in the action, with the anticipated negative kinetic term for the
conformal field $\phi$. In the
perturbative theory, this is not a real problem since the conformal
term can be isolated explicitly and eliminated. However, the 
ensuing unboundedness of the action spells potential trouble
for any non-perturbative geometric path integral (that is either
Euclidean from the outset, or has been Euclideanized by a suitable
Wick rotation), since the Euclidean weight factors
exp($-S)=$exp($\dot\phi^2+\dots )$ can become arbitrarily large.  
We will see that this problem arises in our approach too,
and how it is resolved non-perturbatively.

First to some basics of Lorentzian dynamical triangulations in
three dimensions. The construction of space-time manifolds is completely
analogous to the 2d case. Slices of constant integer $t$ are now
two-dimensional space-like, equilateral triangulations of a given,
fixed topology ${}^{(2)}\Sigma$, and
time-like edges interpolate between adjacent slices $t$ and $t+1$.
The building blocks are given by two types of tetrahedra:
one of them has three space-like and three time-like edges, and shares
its space-like face with a slice $t=$const, the other has four
time-like and two space-like edges, the latter belonging to two
distinct adjacent spatial slices (Fig.\ref{tetra}). We often denote the different
tetrahedral types by the numbers of vertices $(n,m)$ they have in
common with two subsequent slices, which in three dimensions
can take the values (3,1) (together with its time inverse (1,3)) and 
(2,2). Within a given sandwich $\Delta t =1$, a (2,2)-tetrahedron
can be glued to other (2,2)'s, as well as to (3,1)- and (1,3)-tetrahedra,
but a (1,3) can never be glued directly to a (3,1), since their
triangular faces do not match.

\begin{figure}[h]
\vspace{0.5cm}
\centerline{\scalebox{0.6}{\rotatebox{0}
{\includegraphics{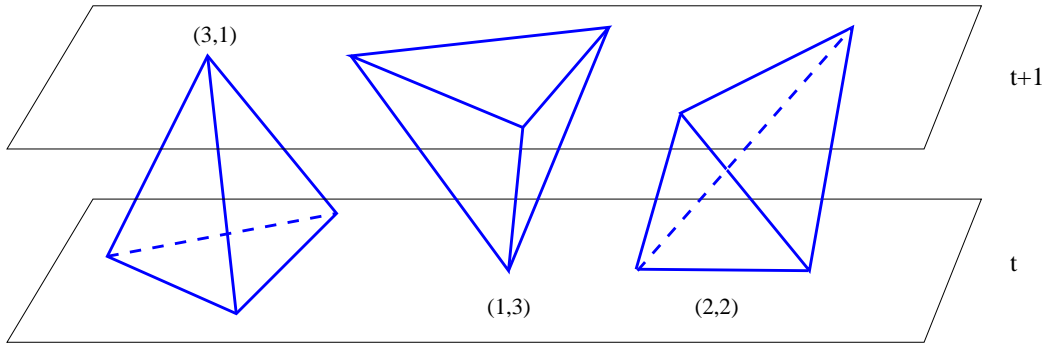}}}}
\vspace{0.5cm}
\caption[tetra]{The three types of tetrahedral building blocks
used in 3d Lorentzian gravity.}
\label{tetra}
\end{figure}

The simplicial action after the Wick rotation reads
\begin{equation}
S= -\kappa_1 N_1(T) +\kappa_3 N_3(T)\equiv 
N_3(T) \Bigl(-\kappa_1 \frac{N_1(T)}{N_3(T)} +\kappa_3\Bigr),
\label{3dact}
\end{equation}
where the latter form is useful in the discussion of Monte-Carlo
simulations, which are usually performed at (approximately)
constant volume.
The phase structure of  the 3d model with spherical spatial topology,
 ${}^{(2)}\Sigma =S^2$, has been determined with the help of
numerical simulations \cite{ajl2}. As expected, there is a critical
line $\kappa_3^{\rm crit}(\kappa_1)$. After fine-tuning to this line,
there is no further phase transition\footnote{The first simulations did report a
first-order transition at  large $\kappa_1$, but this was presumably a
numerical artefact; upon slightly generalizing the class of allowed
geometries, this transition has now disappeared \cite{ajl4}.} along it as a 
function of the inverse Newton coupling $\kappa_1$.

Where is our conformal-mode problem? If we keep the total volume
$N_3$ fixed, the Euclidean action is not actually unbounded, but
because of the nature of our regularization restricted by the range of
the ``order parameter" $\xi:= N_1/N_3$ which kinematically can
only take values in the interval $[1,5/4]$ \cite{d3d4}. This by no means implies
we have removed the problem by hand. Firstly, one can explicitly
identify configurations which minimize the action (\ref{3dact})
and, secondly, the unboundedness could well be recovered upon 
taking the
continuum limit. However, what happens dynamically is that even
in the continuum limit (as far as can be deduced from the simulations
\cite{ajl2,adjl}),
$\xi$ stays bounded away from its ``conformal maximum", which means
that the quantum theory of Lorentzian 3d gravity is {\it not} dominated
by the dynamics of the conformal mode. Configurations with minimal
action exist, but they are entropically suppressed. This is clearly a
non-perturbative effect which involves not just the action, but also
the ``measure" of the path integral. A similar argument of a 
non-perturbative cancellation between certain Faddeev-Popov 
determinants and the conformal divergence can be made in
a gauge-fixed continuum computation \cite{conformal}.\footnote{Of 
course, since the continuum path integral cannot really be {\it done}
(strictly speaking, not even in two dimensions), 
the cancellation argument has to rely on certain (plausible) assumptions
about the behaviour of the path integral under renormalization.}

This result is reassuring, because it shows that 
(Euclideanized) path integrals are not doomed to fail, if only they
are set up properly and non-perturbatively. It also agrees with the
expectation one has from canonical treatments of the theory 
where it is obvious that the conformal mode is not a
propagating degree of freedom. 

\begin{figure}[t]
\vspace{-3.0cm}
\centerline{\scalebox{0.6}{\rotatebox{0}{\includegraphics{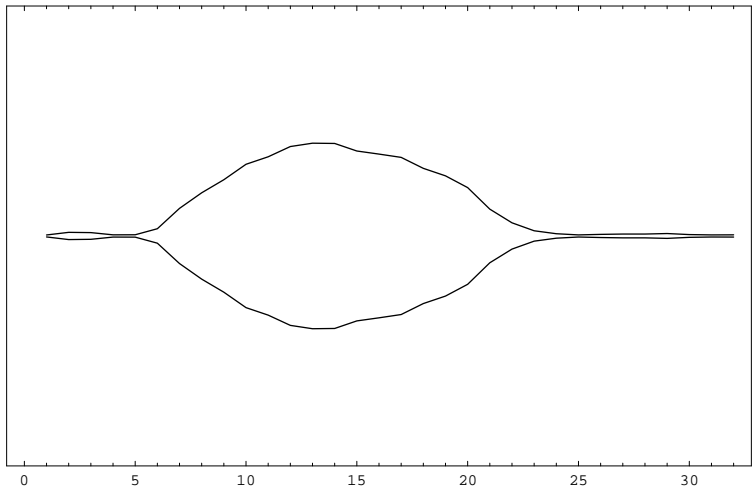}}}}
\vspace{-3.0cm}
\caption[universe]{A typical three-dimensional universe, represented 
as a distribution of two-volumes 
$N_{2}(t)$ of spatial slices at proper times $t\in [0,32]$, at $k_0=5.0$.}
\label{universe}
\end{figure} 

What can we say about the quantum dynamics of 3d Lorentzian
gravity and the geometry of its ground state? Fig.\ref{universe} shows a 
snapshot of a typical ``universe" produced by the Monte-Carlo
simulations. The only variable plotted as a function of the discrete
time $t$ is the two-volume of a spatial slice. What has been determined
are the macroscopic scaling properties of this universe; they are
in agreement with those of a genuine three-dimensional compact
space-time, its time extent scaling $\propto N_3^{1/3}$ and its
spatial volume $\propto N_3^{2/3}$. 

Current efforts are directed at trying to analyze the detailed
microscopic geometric properties of the quantum universe,
its effective quantum Hamiltonian, and at gaining an explicit 
analytic understanding of the conformal-factor cancellation. 
One important question is how
exactly the conformal mode decouples from a propagator
like $G(g^{({\rm in})},g^{({\rm out})})$, although it appears
among the labels parametrizing the in- and out-geometries
$g$. One does not in general expect to be able to make much
progress in solving a three-dimensional statistical model
analytically. However, we anticipate some simplifying features
in the case of pure three-dimensional gravity, which is known to
describe the dynamics of a finite number of physical 
parameters only. 

There are two main strands of investigation, one for space-times
${\bf R}\times S^2$ and using matrix model techniques, and
 the other for space-times ${\bf R}\times T^2$ with flat toroidal
spatial slices. An observation that is being used in both is the
fact that the combinatorics of the transfer matrix, crucial to the
solution of the full problem, is encoded in a {\it two}-dimensional
graph. The transfer matrix $\hat T$, defined in analogy
with (\ref{prop1s}), describes all possible transitions from
one spatial 2d triangulation to the next. Such a transition is
nothing but a three-dimensional sandwich geometry $[t,t+1]$, 
and is completely characterized by the two-dimensional
pattern that emerges when one intersects this geometry at
the intermediate time $t+1/2$. 
One associates with each time-like triangle a coloured
edge where the triangle meets the $(t+1/2)$-surface. A blue
edge belongs to a triangle whose base lies in the triangulation
at time $t$, and a red edge denotes an upside-down triangle with
base at $t+1$. The intersection pattern can therefore be viewed
as a combined tri- and quadrangulation, made out of red triangles,
blue triangles, and squares with alternating red and blue sides. 

Graphs of this type, or equivalently their duals, are also generated by
the large-$N$ limit of a hermitian two-matrix model with
partition function
\begin{equation}
Z(\alpha_1,\alpha_2,\beta)=\int dA_{N\times N}\ dB_{N\times N}\
{\rm e}^{-N\ {\rm Tr}(\frac{1}{2} A^2+\frac{1}{2} B^2-\alpha_1 A^3
-\alpha_2B^3 -\beta ABAB)}.
\label{mamo}
\end{equation} 
The cubic and quartic interaction terms in the exponent correspond
to the tri- and four-valent intersections of the dual bi-coloured 
spherical graph characterizing a piece of space-time. In fact, as
was shown in \cite{ajlv}, the matrix model gives an embedding 
of the gravitational model we are after, since it generates {\it
more} graphs than those corresponding to regular three-dimensional
geometries. Interestingly, from a geometric point of view these can
be interpreted as wormhole configurations.  Some explicit examples 
are shown in Fig.\ref{worms}; the graphs consist of squares since
they are taken from a ``pyramid" variant of three-dimensional
gravity, cf. footnote 16. Blue and red edges are in these pictures
represented by solid and dashed lines.

\begin{figure}[t]
\vspace{0.5cm}
\centerline{\scalebox{0.5}{\rotatebox{0}
{\includegraphics{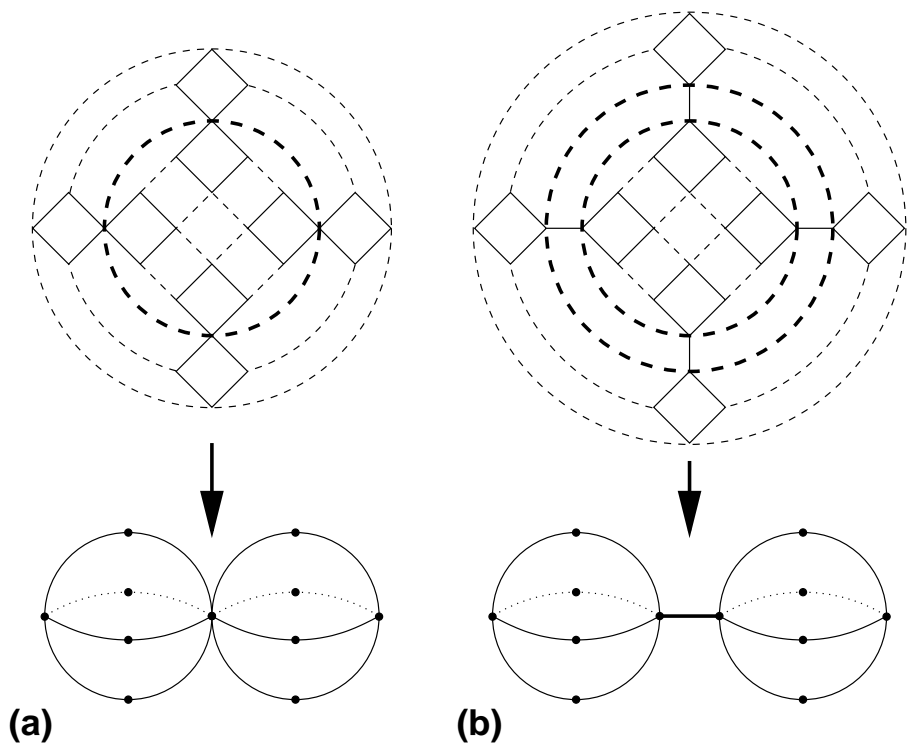}}}}
\caption[worms]{Examples of quadrangulations at $t+1/2$ corresponding 
to wormholes at time $t$.
Shrinking the dashed links to zero, one obtains the two-geometries 
at the bottom. The thick dashed lines at the 
top are contracted to points where wormholes begin or end.}
\vspace{0.5cm}
\label{worms}
\end{figure}

The matrix model has
been solved analytically for the diagonal case $\alpha_1=\alpha_2$ \cite{kz},
and its second-order phase transition separates the phase where
wormholes are rare from that where they are abundant.\footnote{More 
precisely, these results apply to a variant of (\ref{mamo})
where the cubic terms $A^3$ and $B^3$ have been replaced by
quartic terms $A^4$ and $B^4$. Geometrically, this corresponds
to using pyramids instead of the tetrahedral building blocks, a
difference that is unlikely to affect the continuum theory.}  One
therefore concludes that Lorentzian gravity as given by dynamical
triangulations should correspond to the former. 

It turns out that to extract information about the quantum Hamiltonian
of the system, one must consider the off-diagonal case where
the two $\alpha$-couplings are different. Only in that
case can one distinguish which part of the intersection
graph comes from ``below" (time $t$) and which from ``above" 
(time $t+1$). The colouring of the two-dimensional graph is
really the memory of the original three-dimensional nature of the
problem. It turns out that even for $\alpha$'s which differ only infinitesimally,
this is a highly non-trivial problem. Making a natural ansatz for the
analytic structure of the eigenvalue densities that appear in the
partition function, a consistent set of equations has now been
found, which will hopefully yield more details about the effective
Hamiltonian of the quantum system \cite{ajjlv}. Since there are no non-trivial
Teichm\"uller parameters in the sphere case, what one might expect 
on dimensional grounds is a
differential operator in the two-volume $V_2$ of the kind \cite{ajl4}
\begin{equation}
\hat H= -c_1 G_N V_2\frac{d^2}{dV_2^2}-c_2 \Lambda V_2,
\label{3dham}
\end{equation}
where the $c_i$ are numerical constants.

A second direction of attack are {\it cosmological models} of
3d gravity. They are symmetry-reduced in the sense that only 
a restricted class of spatial geometries is allowed at integer values
of $t$, and also additional conditions may be imposed on the
interpolating three-dimensional Lorentzian geometries. 
All models studied so far have flat tori as their spatial slices,
the simplest case with a non-trivial physical configuration
space, spanned by two real Teichm\"uller parameters (apart
from the two-volume of the spatial slices). Flat two-dimensional
tori can be obtained by suitably identifying the boundaries of a
piece of the triangulated plane. Since we are working with
equilateral triangles, this amounts to a piece of regular
triangulation where exactly six triangles meet at every (interior)
vertex point. 

Even if the spatial slices have been chosen as spaces of constant
curvature, this still leaves a number of possibilities of how the
space-time in between can be filled in. One extreme choice would be
to allow any intermediate three-geometry. By this we would 
probably not gain much in terms of simplifying the model, which 
obviously is a major motivation behind going ``cosmological". 
By contrast, the first model studied had very simple
interpolating geometries. The most transparent realization of this
model is in terms of (4,1)- and (1,4)-pyramids rather than the 
(3,1)- and (1,3)-tetrahedra (a modification we already encoutered
in the discussion of the matrix model), so that the
spatial slices at integer-$t$ are regular square lattices \cite{dehne}.
The corresponding 2d building blocks of the intersection graph
at half-integer $t$ are now blue squares, red squares and -- as before --
red-and-blue squares. If the (cut-open) tori at times $t_1=t$ 
and $t_2=t+1$ consist of $l_i$ columns and $m_i$ rows, $i=1,2$,
any allowed intersection pattern is a rectangle of size
$(l_1+l_2)\times(m_1+m_2)$. An example is shown in Fig.\ref{cmodel}.
The trouble with this simple model is that it does not have
enough entropy: the number of possible interpolating
sandwiches between two neighbouring spatial slices is given by
\begin{equation}
{\rm entropy}\propto \Biggl({l_1+l_2\atop l_1}\Biggr)
\Biggl({m_1+m_2\atop m_1}\Biggr),
\label{3dentropy}
\end{equation}
which is roughly speaking the square of the entropy of the 
two-dimensional Lorentzian model, cf. equation (\ref{prop1s}).
This is not enough in the sense that the number of
``microstates" in a piece of space-time $\Delta t=1$  scales
asymptotically only with the linear size of the tori, ie.
like exp($c\cdot$length). Such a behaviour cannot ``compete"
with the exponential damping exp($c'\cdot$area) coming from
the cosmological term in the action. Thus, the only space-times
that will not be exponentially damped in the continuum limit
will be those whose spatial slices are essentially one-dimensional.
This clearly is a limit that has nothing to do with the
description of 3d quantum geometries we are after. In particular, 
the model is unsuitable for studying the conformal-mode cancellation.

\begin{figure}[t]
\centerline{\scalebox{0.5}{\rotatebox{0}
{\includegraphics{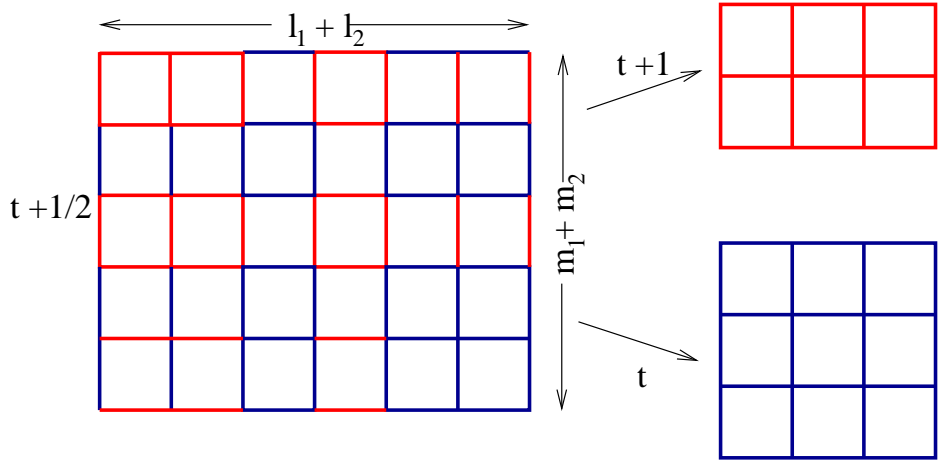}}}}
\caption[cmodel]{The cosmological ``pyramid model" has regular
slices at both integer and half-integer times.}
\label{cmodel}
\end{figure}

I have included a discussion of this model
because it suggests a potential problem for the path integral in
models that impose severe symmetry constraints {\it before}
quantization. Prime examples of this are continuum 
mini-superspace models
with only a finite number of dynamical degrees of
freedom, whose path integral formulations are 
riddled with difficulties. 
Lorentzian dynamically triangulated models are more flexible
concerning the imposition of such constraints. 

The next 
cosmological model I will consider has also flat tori at
integer-$t$, but allows for more general geometries in
between the slices. As a consequence, it does not suffer from the
problem described above. The easiest way of describing
the geometry of this so-called {\it hexagon model} is by
specifying the intersection patterns at half-integer $t$.
One such pattern can be thought of as a tiling of a regular piece of
a flat equilateral triangulation with three types of coloured rhombi.
The colouring of the rhombi again encodes the orientation
in three dimensions of the associated tetrahedral building
block. A blue rhombus stands for a pair of (3,1)-tetrahedra,
glued together along a common time-like face,  a red rhombus
for a pair of (1,3)-tetrahedra, and the rhombus with alternating
blue and red sides is a (distorted) representation of a 
(2,2)-tetrahedron. Opposite sides of the regular triangular
``background lattice" are to be identified to create the topology
of a two-torus. The beautiful feature of this model is the fact
that any complete tiling of this lattice by matching rhombic 
tiles {\it automatically} gives rise to flat two-tori on the two
spatial boundaries of the associated sandwich $[t,t+1]$
\cite{hexagon}.

After the Wick rotation, the one-step propagator of this model
can be written as 
\begin{equation}
G(g^{(1)},g^{(2)};\Delta t=1)\equiv \langle g^{(2)}|\hat T|
g^{(1)}\rangle={\cal C}(g^{(1)},g^{(2)})\ {\rm e}^{-S(g^{(1)},g^{(2)})}.
\label{3dprop}
\end{equation}
We note here a distinguishing property of the hexagon model,
namely, a factorization of $G$ into a Boltzmann weight  exp($-S$)
and a combinatorial term $\cal C$ which counts the number of 
distinct sandwich geometries with fixed toroidal boundaries
$g^{(1)}$ and $g^{(2)}$, both of which depend on the boundary
data only, and not on the details of the three-dimensional 
triangulation of its interior. The leading asymptotics of the 
entropy term is determined by the combinatorics of a model
of so-called {\it vicious walkers}. The walkers are usually
represented by an ensemble of paths that move up a tilted
square lattice, taking steps either diagonally to the left or
to the right, in such a way that at most one path passes through
any one lattice vertex. 

\begin{figure}[t]
\vspace{.5cm}
\centerline{\scalebox{0.45}{\rotatebox{0}
{\includegraphics{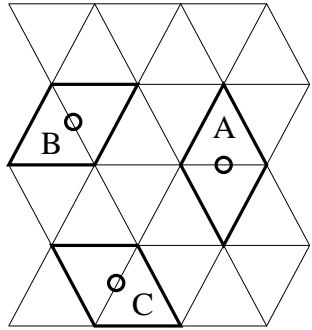}}}}
\caption[abc]{A rhombus can be put onto the triangular background
lattice with three different orientations, A, B or C.}
\vspace{.5cm}
\label{abc}
\end{figure}

\begin{figure}[h]
\vspace{.5cm}
\centerline{\scalebox{0.45}{\rotatebox{0}
{\includegraphics{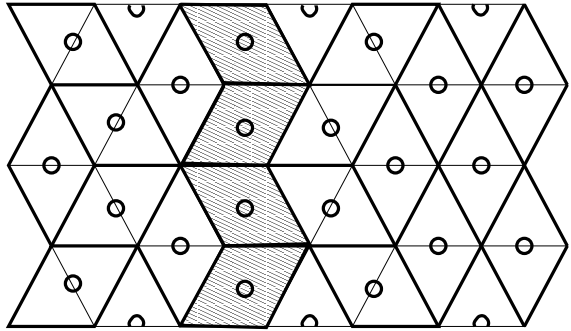}}}}
\caption[tiling]{An example of a periodic tiling of the
triangular background lattice. The shaded
region is a B-C-path with winding number (0,1).}
\vspace{.5cm}
\label{tiling}
\end{figure}

The paths of the hexagon model are sequences of rhombi that
have been put down on the background lattice so they lie on
one of their sides (types B and C in Fig.\ref{abc}). Because of the
toroidal boundary conditions, such B-C-paths wind around the 
background lattice in the ``vertical direction" (on figures such
as Fig.\ref{tiling}), which for the purposes of solving the 2d statistical
model of vicious walkers we may think of as the time
direction. The transfer matrix of this model can be diagonalized
explicitly. Let us denote the number of vicious-walker paths by
$w/2$, the width of the background lattice by $l+w$ and its 
height (in time direction) by $m$, all in lattice units. It turns out
that for the simplest version of the model we can set $m=l$
without loss of generality. We are now interested in the
number ${\cal N}(l,w)$ which solves the following
combinatorial problem: 

\vspace{.6cm}
\parbox{13.4cm}{Given two even integers $l$ and
$w$, how many ways ${\cal N}(l,w)$ are there of drawing $w/2$ 
non-intersecting 
paths of winding number $(0,1)$ (in the horizontal and vertical
direction) onto a tilted square lattice of width $l+w$ and height 
$l$, with periodic boundary conditions in both directions?}
\vspace{.6cm}

\noindent Denoting by $\vec\lambda=(\lambda_1,\dots,\lambda_{w/2})$,
$\lambda_i\in\{0,2,4,\dots,l+w\equiv 0\}$, the vector of
positions of the vicious walkers along the horizontal axis, the
eigenvectors of the transfer matrix have the form
\begin{equation}
\Psi (\vec\lambda)=\frac{1}{\sqrt{\frac{w}{2}!} }\det [z_j^{\lambda_i}],
\;\;\;\;\; 1\leq i,j\leq \frac{w}{2},
\label{eigenvec}
\end{equation}
where the complex numbers $z_j$ are given by
\begin{equation}
z_j={\rm e}^{i\pi \frac{k_j}{l+w}}\  {\rm e}^{i\pi \frac{w-2}{l+w}},\;\;\;\;\;
0\leq k_1<k_2 \dots < k_{w/2}\leq \frac{l+w}{2}-1.
\label{zdefine}
\end{equation}
This result can be understood by observing that for
a single walker in the same representation, taking a step to the
right (left) is represented by a multiplication (division) by $z$, that is,
\begin{equation}
\Psi(\lambda)=z^\lambda \;\;\;\Longrightarrow \;\;\;
z\Psi(\lambda) =z^{\lambda +1}\equiv \Psi(\lambda +1).
\label{zaction}
\end{equation}
The expression (\ref{eigenvec}) is an appropriately antisymmetrized and
normalized version for the case of several walkers. In this
representation, the transfer matrix\footnote{This is the transfer matrix
corresponding to a ``double step" in time; a single step would lead to a
position vector with odd $\lambda_i$'s.} takes the form
\begin{equation}
\hat T_{\rm VW}=\prod_{i=1}^{w/2} \bigl( \frac{1}{z_i} +2 +z_i\bigr).
\label{vwtrans}
\end{equation}
The final result in the limit as both $l,w\rightarrow\infty$, with a
fixed ratio $\alpha :=\frac{w}{l+w}$, is to leading order given by
\begin{equation}
{\cal N}(l,w)=C(\alpha)^{\frac{lw}{2}},\;\;\;\;\;
C(\alpha) =\exp \Biggl[ \frac{2}{\alpha} \int_0^{\alpha/2} dy\ 
\log(2 \cos\pi y) \Biggr].
\label{nfinal}
\end{equation}
This shows that the hexagon model has indeed enough entropy,
since the number of possible intermediate geometries scales
exponentially with the area, and not just with the linear dimension
of the tori involved. 

Another attractive feature of the model is that the Teichm\"uller
parameters $\tau(t)=\tau_1(t)+i\tau_2(t)$ of the spatial tori at
time $t$ can be written explicitly as functions of the discrete
variables describing the Lorentzian simplicial space-time. 
It turns out that the real parameter $\tau_1$ is not dynamical,
so that the wave functions of the model are labelled by
just two numbers, the two-volume $v(t)$ and 
$\tau_2(t)$.\footnote{The model can be generalized to have
non-trivial $\tau_1$ by allowing for B-C-paths with higher
winding numbers \cite{hexanew}.}
Expanding the euclideanized action for small $\Delta t=a$,
one finds
\begin{equation}
S=\tilde\lambda v-\tilde k a^2v\Bigl(  \Bigl(\frac{\dot v}{v}\Bigr)^2-
\Bigl(\frac{\dot \tau_2}{\tau_2}\Bigr)^2 \Bigr)+\dots,
\label{3dexpand}
\end{equation}
where $\tilde\lambda$ and $\tilde k$ are proportional to the bare
cosmological and inverse Newton's constants. This has the expected
modular-invariant form, with a standard kinetic term for
$\tau_2$, and one with the wrong sign for the area $v$. Of course,
this is our old friend, the (global) conformal mode!

What we are after is the ``effective action", containing contributions
from both (\ref{3dexpand}) and the state counting, namely,
\begin{equation}
S^{\rm eff}:= S-\log ({\rm entropy})=v(\tilde\lambda -C)+\; ???
\label{quest}
\end{equation}
In order to say anything about the cancellation or otherwise of
the conformal divergence, we need more than just the 
leading-order term (\ref{nfinal}) of the entropy of the hexagon
model. Unlike the exponential term, these subleading terms are 
sensitive to the colouring of the intersection graph, and efforts
are under way to solve the corresponding vicious-walker problem
\cite{hexanew}. 

\subsection{Beyond three dimensions}

As already mentioned earlier, there is nothing much to
report at this stage on the nature of the continuum limit in the 
physical case
of four dimensions. The first Monte-Carlo simulations are just
being set up, but any conclusive statements are likely to involve
a combination of analytical and numerical arguments. Also it
should be kept in mind that, unlike in previous simulations of
four-dimensional {\it Euclidean} dynamical triangulations,
the space-times involved here are not isotropic. 
Measurements of two-point functions, say, will be sensitive
to whether the distances are time- or space-like, and therefore
more computing power will be necessary to achieve a statistics
comparable to the Euclidean case. 

One way of making progress in four dimensions will be by
studying geometries with special symmetries, along the lines
of the 3d cosmological models discussed above. It should
be noted that popular symmetry reductions, such as
spherical or cylindrical symmetry, cannot be implemented
{\it exactly} because of the nature of our discretization. 
They can at best be realized approximately, which in view of
the results of the previous subsection may be a good thing
since it will ensure that a sufficient number of microstates
contributes to the state sum. An important application
in this context is the construction of a
path integral for spherical black hole configurations. 
Already the formulation of the problem has a number of 
challenging aspects, for example, the inclusion of non-trivial 
boundaries, an explicit realization of the (near-)spherical 
symmetry, and of a ``horizon finder", some of which
have been addressed and solved in 
\cite{kappel,dittrich}. It will be extremely
interesting to see what Lorentzian dynamical triangulations have 
to say about the famous thermodynamic properties of
quantum black holes from a non-perturbative point
of view. These questions are currently under study.

\section{Brief conclusion}\label{conclusion}

As we have seen, the method of {\it Lorentzian dynamical triangulations}
constitutes a well-defined regularized framework for constructing
non-perturbative theories of quantum gravity. Technically, they
can be characterized as regularized sums over simplicial
random geometries with a time arrow and certain
causality properties. 
In dimension $d<4$, interesting continuum limits have been
shown to exist. Their geometric properties have been explored,
almost exhaustively in two, and partly in three
dimensions. Both are examples of Lorentzian quantum gravitational
theories which as continuum theories are {\it inequivalent} to their
Euclidean counterparts, and the relation between the two is
{\it not} that of some simple analytic continuation of the form
$t\mapsto it$. The origin of the discrepancy between quantum
gravity with Euclidean and Lorentzian signature lies in the absence
of causality-violating branching points for geometries in the 
latter. Since in 
dimension $d\geq 3$, the approach of {\it Euclidean} dynamical
triangulations seems to have serious problems, I am greatly
encouraged by the fact that the 3d Lorentzian model is better
behaved. Of course, it still needs to be 
verified explicitly that the imposition of causality conditions is indeed
the correct remedy to cure the {\it four}-dimensional theory of its
apparent diseases. One step in that direction will be to
show that the non-perturbative cancellation mechanism for the
conformal divergence is also present in $d=4$.

Two warnings may be in order at this point. Firstly, there is a priori 
nothing {\it discrete}
about the quantum gravitational theories this method produces.
Its ``discreteness" refers merely to the intermediate regularization
that was chosen to make the non-perturbative path sums 
converge.\footnote{It should maybe be emphasized that there are 
precious few methods 
around that get even that far, and possess a coordinate-invariant cutoff.} 
In particular, there is nothing in the construction 
suggesting the presence of any kind of ``fundamental discreteness",
as has been found in canonical models of four-dimensional quantum gravity
\cite{rovellismolin,loll1,loll2}. 
Secondly, one should refrain from trying to interpret the discrete
expressions of the regularized model as some kind of
approximation of the ``real" quantum theory {\it before} one has shown
the existence of a continuum limit which (at least in dimension four) 
is an interacting theory of geometric degrees of freedom.

In conclusion, I have described here a possible path for constructing a
non-pertur\-ba\-tive quantum theory of gravity, by applying standard
tools from both quantum field theory and the theory of critical
phenomena to theories of fluctuating geometry. Investigation of
the continuum theories in two and three space-time dimensions has already
led to exciting new insights into the relation between the Lorentzian and
Euclidean quantum theories, and ways of understanding and 
resolving the conformal
sickness of gravitational path integrals, as well as bringing in new
tools from combinatorics and statistical mechanics. I hope this has
convinced you that the method of Lorentzian dynamical triangulations
stands a good chance of throwing some light on the ever-elusive
quantization of general relativity!

\vspace{1cm}
\noindent {\it Acknowledgements.} I would like to thank
D.\ Giulini, C.\ Kiefer and C.\ L\"ammerzahl for organizing an
interesting and inspiring meeting at Bad Honnef, and for inviting
me to speak there. Support through the
EU network on ``Discrete Random Geometry'', grant HPRN-CT-1999-00161, 
and the ESF network no.82 on ``Geometry and Disorder'' is also
gratefully acknowledged.

\end{document}